\documentclass{article}
\usepackage{arxiv}
\usepackage[utf8]{inputenc} 
\usepackage[T1]{fontenc}    
\usepackage{hyperref}       
\usepackage{url}            
\usepackage{booktabs}       
\usepackage{amsfonts}       
\usepackage{nicefrac}       
\usepackage{microtype}      
\usepackage{lipsum}		
\usepackage{graphicx}
\usepackage{natbib}
\usepackage{doi}
\usepackage{lscape}

\newcommand{\invcm}{cm$^{-1}$}
\newcommand{\xsec}{cm$^2$ molecule$^{-1}$}

\title{\textsc{ExoMol} photodissociation cross sections I: HCl and HF}

\author{Marco Pezzella, Jonathan Tennyson, Sergei N. Yurchenko \\
	Department of Physics \& Astronomy,\\
	University College London,\\
	London WC1E~6BT, UK\\
}

\date{}

\renewcommand{\headeright}{ }
\renewcommand{\undertitle}{ }
\renewcommand{\shorttitle}{\textsc{ExoMol} photodissociation I: HCl and HF}

\hypersetup{
pdftitle={\textsc{ExoMol} photodissociation I: HCl and HF},
pdfsubject={q-bio.NC, q-bio.QM},
pdfauthor={Marco Pezzella, Jonathan Tennyson, Sergei N. Yurchenko},
}
\begin{document}
\maketitle

\begin{abstract}
Photon initiated chemistry, \textit{i.e.} the interaction of light with chemical species, is a key factor in the evolution of the atmosphere of exoplanets. For planets orbiting stars in UV-rich environments, photodissociation induced by high energy photons dominates the atmosphere composition and dynamics. The rate of photodissociation can be highly dependent on atmospheric temperature, as  increased temperature leads to  increased  population of  vibrational excited states and the consequent lowering of the photodissociation threshold. 
This paper inaugurates a  new series of papers  presenting computed temperature-dependent photodissociation cross sections with  rates generated  for different stellar fields. Cross sections  calculations are performed by solving the time-independent Schr\"{o}dinger equation for each electronic state involved in the process. Here photodissociation cross sections for hydrogen chloride and hydrogen fluoride are
computed for a grid of 34 temperatures between 0 and 10~000 K. Use of different radiation fields shows that for the Sun and cooler stars the photodissociation rate can increase exponentially for molecular temperatures above 1000 K; conversely the photodissociation rates in UV rich fields instead are almost insensitive to the temperature of the molecule. Furthermore, these rates show extreme sensitivity to the radiation model used for cool stars, suggesting
that further work on these may be required. The provision of an ExoMol database of cross sections  is discussed.  \end{abstract}


\section{Introduction}

The interaction of photons with molecules, known as photochemistry, plays an important role in the atmospheric composition and dynamics of planets and exoplanets. It is especially important for all the exoplanets that orbit near their host star, as those planets experience strong UV-rich stellar fields \citep{16VeRoCa,16MaAgMo,19BaWoKo,19FlGuHe,20LeWaMa,22TeKeBa.fields}. Measurements of photodissociation cross sections  have been performed mostly at low temperatures, but, even when high temperature measurements are available \citep{16VeRoCa}, they struggle to reach temperatures of interest for planetary atmospheres ($T>1000$ K). The existing models and databases \citep{Leiden,UVVis+} are only adequate for cold molecules in the interstellar medium, where molecules are in their vibrational ground states \citep{20VaBeGr}. There is a long history of theoretical photodissociation studies \citep{93Schinke}, but only recently was the importance of the temperature effects demonstrated \citep{16Greben}. Those effects are important for modelling non-local thermodynamic effects (non-LTE), with the photodissociation being one of the major driving forces of the non-LTE effects \citep{21ClYu}, along with radiative, (photo-)chemical, energy transfer and other processes \citep{01LoTaxx}.

The \textsc{ExoMol} database \citep{jt631,jt810} was designed to produce comprehensive line lists of hot bound-bound transitions for molecules that are important in exoplanet atmospheres \citep{jt528,jt810}. In our previous work \citep{jt840} we showed how the programs \textsc{Duo} \citep{jt609}, originally developed to solve the bound-bound nuclear motion problem for diatomics using a grid-based variational solution of the Schr\"{o}dinger equation including different types of couplings and crossings, and \textsc{ExoCross} \citep{jt708} can be used to produce photodissociation cross sections by averaging the results of \textsc{Duo} calculations obtained using different radial grids and smoothing the results with an appropriate Gaussian function.  With this work we  start a new branch of the \textsc{ExoMol} project that aims to provide temperature-dependent photodissociation data (cross sections and rates) for molecules found in exoplanetary atmospheres and elsewhere. 
The first two molecules studied are HCl and HF. The  A\,$^1\Pi \leftarrow$X\,$^1\Sigma^+$ photodissociation transitions of these molecules formed part of the test cases used to shape our methodology \citep{jt840};
here we extend our model to more states covering both direct and indirect photodissociations and consider the effects of temperature on the molecules.  

Chlorine has a relative abundance with respect to hydrogen of $3\times 10^{-7}$ \citep{09AsGrSa}; it has two stable isotopologues ($^{35}$Cl, $^{37}$Cl), with a terrestrial isotope ratio close to 3:1. HCl is the only chlorine-bearing molecule observed in the interstellar media. The molecule has been detected in the circumstellar envelope of the carbon rich star IRC+10216, with a molecular abundance with respect to H$_2$ of $6 \times 10^{-7}$ \citep{10CeDeBa.HCl}, in molecular clouds as OMC-1 \citep{85BaKePh.HCl,95ScPhWa.HCl}, Sagittarius B2 \citep{95ZmBlCa.HCl}, $\rho$ Ophiuci \citep{95FeCaDi.HCl}, and W31C \citep{13MoLiRo.HCl} . HCl has been used by \citet{18MaPi.HCl} for determining the isotope ratio $^{35}$Cl/$^{37}$Cl in six M giant stars, finding an average value of   $\frac{^{35}{\rm Cl}}{^{37}{\rm Cl}} = 2.66 \pm 0.58$. The molecule is expected to be found in low quantities on Jupiter's atmosphere \citep{14TeShFl.HCl}.  There is only one stable isotope of fluorine with an abundance relative to hydrogen of $3.6 \times 10^{-8}$ \citep{09AsGrSa}. Hydrogen fluoride has been identified by \citet{97NeZmSC.HF} in the interstellar medium (for example in the W31C \citep{10NeSoPh.HF}, W49N and W51 \citep{10SoNePh.HF}) and in cool stellar atmospheres of AGB stars by \citet{08UtArLe.HF}, and later was identified in two nearby galaxies (NGC 253 and NGC 4945) with an estimated abundance with respect to H$_2$  of $6 \times 10^{-9}$ \citep{14MoLoFa.HF}. Developments in the observations of HCl and HF, and their application in modelling approaches are found in \citet{16GeNeGo.hydrides}.

The paper is structured as follows: the method section describes the methodology used for computing the photodissociation cross sections and rates, giving the sources of our potential energy curves and transition dipole moments. The following two sections present our results and  section 5 briefly outlines the upcoming database. Finally, section 6 gives our conclusions. 

\section{Methods}
\subsection{Potential energy  and transition dipole moment curves}

Although the ExoMol data base provides line lists for both HCl and HF, these were actually taken from the work of \citet{13LiGoLe.HCl,13LiGoHa.HCl}, as incorporated in HITRAN \citep{jt691},
which provides  pure-rotational and ro-vibrational transition data over an extended temperature energy range. This means we had to build spectroscopic models for the rovibronic transitions from scratch for the
current study.

\begin{figure*}
	\includegraphics[width=\columnwidth]{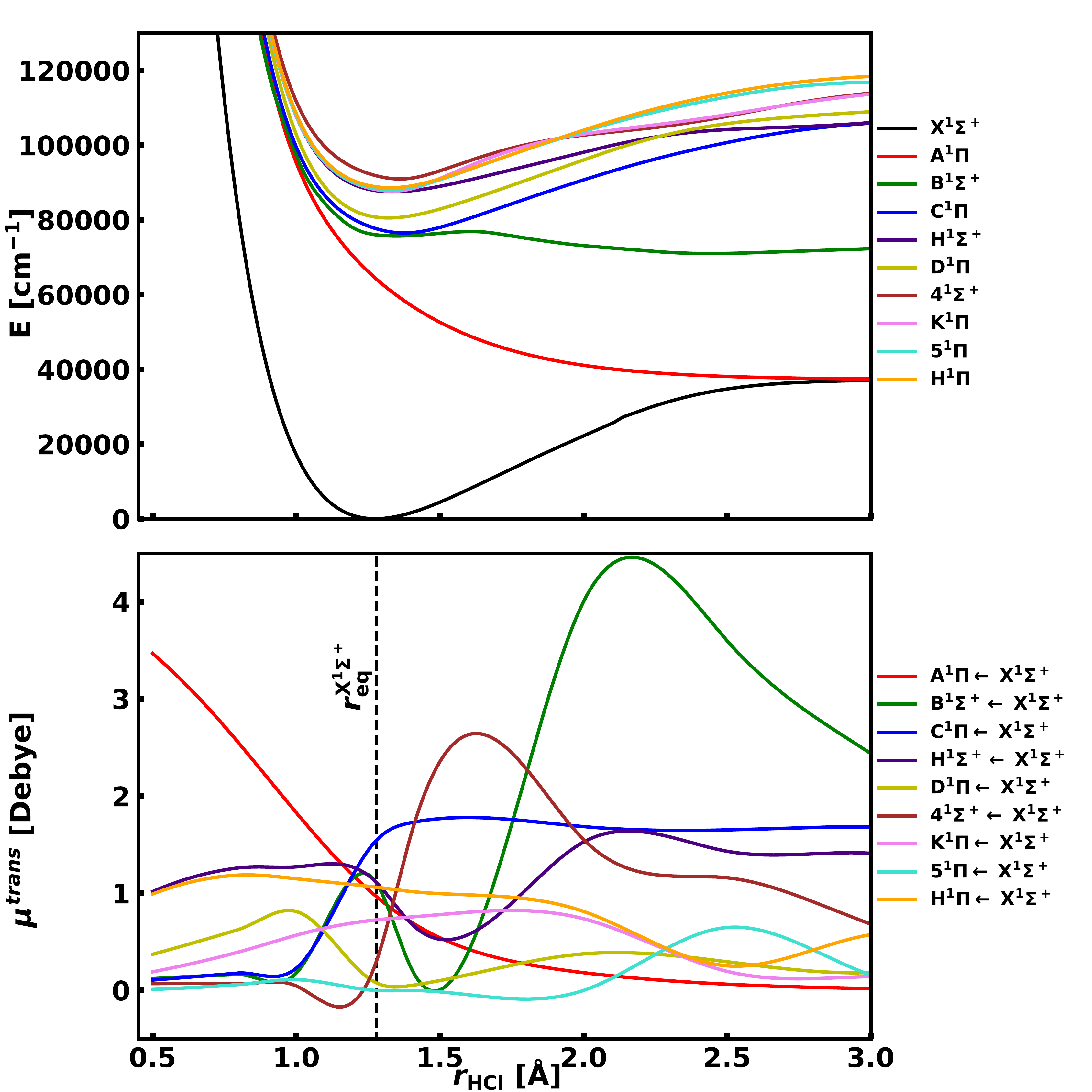}
    \caption{Potential energy curves and transition dipole moments of HCl. PECs of the X\,$^1\Sigma^+$ and  A\,$^1\Pi$ states are taken from \citet{93AlPoDu.HCl},  TDMC of A\,$^1\Pi \leftarrow$X\,$^1\Sigma^+$  is from \citet{86GiBa.HCl}. The other PECs and TDMCs are taken from \citet{12EnSiCa.HCl}.}
    \label{fig:HCl-pecs}
\end{figure*}

We consider 9 electronic states of HCl, whose minima lie below the H ionization limit at approximately 100~000 \invcm, see Fig.~\ref{fig:HCl-pecs}. Potential energy curves (PECs) of the X\,$^1\Sigma^+$ and  A\,$^1\Pi$  states are taken from \citet{93AlPoDu.HCl}, the transition dipole moment curve (TDMC) between those states is from \citet{86GiBa.HCl}. In this case, the choice of using this combination of a PEC and a TDMC from two different sources has been validated by the comparison of our results with the experiments from different sources \citep{jt840}.  All other PECs and TDMCs are taken from \citet{12EnSiCa.HCl}. The potentials are described using the cubic spline interpolation implemented in \textsc{Duo}. The A\,$^1\Pi$ potential is a repulsive state that asymptotically goes to the ground state H($^2S$) + Cl($^2$P) dissociation limit. The B\,$^1\Sigma^+$ state has a double  well structure with the lowest minimum near 2.44 \AA\/  (at 71948.2 \invcm), and a second minimum near 1.35 \AA\/ (at 76643.3 \invcm); the two minima are separated by a barrier of 4695 \invcm. The double well structure of the B\,$^1\Sigma$ state arises from the avoided crossing between two states, identified as E\,$^1\Sigma^+$ and V\,$^1\Sigma^+$, however, for the photodissociation process there is no advantage in treating the two states separately \citep{11LeLiVa.HCl}.  The potential energy curves from \citet{12EnSiCa.HCl} were shifted to improve the agreement with the experimental results from \citet{02ChChBa.HCl} and \citet{06LiZhYu.HCl} for the C\,$^1\Pi$, D\,$^1\Pi$, H$^1\Sigma^+$, K\,$^1\Pi$, M\,$^1\Pi$ states; the B\,$^1\Sigma^+$, 5\,$^1\Pi$ and 4\,$^1\Sigma^+$ are shifted in order to keep constant the energy differences with the surrounding states. The TDMCs are described by interpolation.    

We neglect the coupling contributions from the a\,$^3\Pi\leftarrow$X\,$^1\Sigma^+$, the b\,$^3\Pi \leftarrow$X\,$^1\Sigma^+$ and the t\,$^3\Sigma \leftarrow$X\,$^1\Sigma^+$ bands, as their contributions to the wavefunction are less than 1$\%$ \citep{12EnSiCa.HCl}. The b\,$^3\Pi \leftarrow$X\,$^1\Sigma^+$ band contribution to the photodissociation rates  is estimated to be $1.0 \times 10^{-11}$ s$^{-1}$, less than the $1\%$ of the total photodissociation rate. \citet{70TiGiVa.HI}  observed that the b\,$^3\Pi \leftarrow$X\,$^1\Sigma^+$ cross sections are 40-50 times weaker than the C$^1\Pi \leftarrow$X$^1\Sigma^+$ ones. \citet{08Zhxxxx.HCl} showed that dissociation from the t~$^3\Sigma^+$ state occurs through the spin-orbit coupling with the A~$^1\Pi$ state.

\begin{figure*}
	\includegraphics[width=\columnwidth]{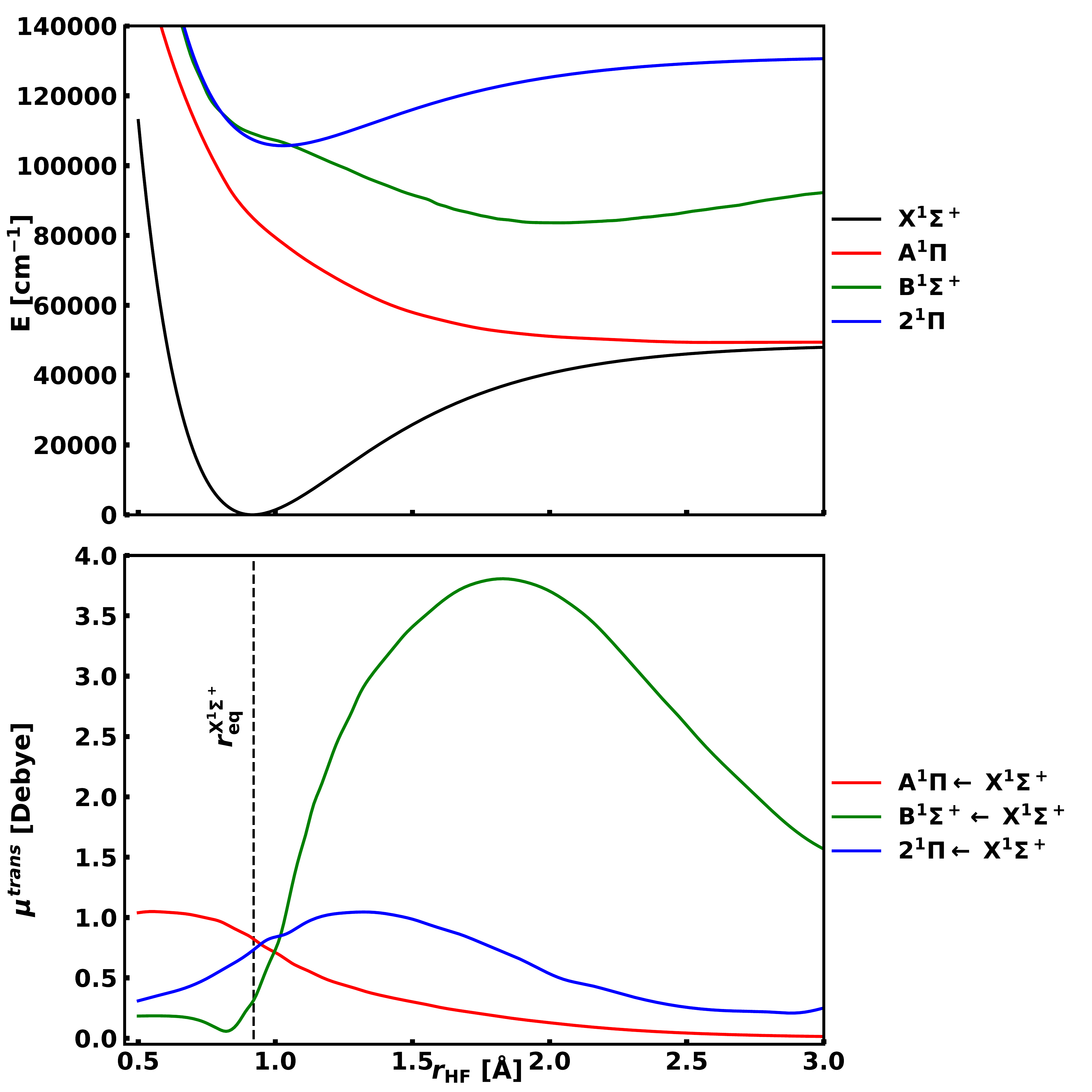}
    \caption{Potential energy  and transition dipole moment curves of HF. PECs of the X\,$^1\Sigma^+$, B\,$^1\Sigma^+$ and C\,$^1\Pi$ states, and the involved transition dipole moments are taken from \citet{21LiSuLi.HF}. PEC of the A\,$^1\Pi$ state and the A\,$^1\Pi \leftarrow$X\,$^1\Sigma^+$ TDMC are from \citet{00BrBa.HF}.}
    \label{fig:HF-pecs}
\end{figure*}

For HF we consider 4 electronic states: X\,$^1\Sigma^+$,  A\,$^1\Pi$, B\,$^1\Sigma^+$, and C\,$^1\Pi$, see Fig.~\ref{fig:HF-pecs}. The PECs of the X\,$^1\Sigma^+$, B\,$^1\Sigma^+$ and C\,$^1\Pi$ states and the transition moments involved  are taken from \citet{21LiSuLi.HF}, the A\,$^1\Pi$ PEC and the A\,$^1\Pi \leftarrow$X\,$^1\Sigma^+$ TDMC are from \citet{00BrBa.HF}. We use  Morse potentials to describe the X\,$^1\Sigma^+$ and C\,$^1\Pi$ PECs. The A\,$^1\Pi$,  B\,$^1\Sigma^+$ PECs and corresponding TDMCs are described using \textsc{Duo}'s cubic spline interpolations. 
The A\,$^1\Pi$ -- X\,$^1\Sigma^+$ band  is the only electronic band below 100~000 \invcm, as the B\,$^1\Sigma^+$ and  C\,$^1\Pi$ states lie above the H$_2$ ionization threshold. 

The assumption that coupling between different electronic states can be neglected without significantly affecting the overall photodissociation cross section process means that separate \textsc{Duo} calculations can
be performed for each excited state. \textsc{Duo} input files for each single electronic excitation are given in the electronic supplementary material: a total of 9 files for HCl and 3  for HF. If one
wishes to consider curve couplings, then multiple electronic states can be considered within the same \textsc{Duo} input structures. 

\subsection{Cross sections calculations}

In this section we briefly describe how  the photodissociation cross sections are computed with a detailed description given by \citet{jt840}. Our method is based on solving the time-independent Schr\"{o}dinger equation using  \textsc{Duo} \citep{jt609} and then post-processing the discretized results using \textsc{ExoCross} \citep{jt708} to give continuous, smooth cross
sections. Our previous numerical experiments showed that use of discretised and smoothed results led to the same photodissociation rates with respect to previously published cross sections \citep{82DiHeDa.HCl,00BrBa.HF,02ChChBa.HCl} when considered with an appropriate (stellar) radiation field; here rates are generated using the smoothed cross sections.

The photodissociation cross sections for each electronic states are evaluated for 34 temperatures between $T=0$ K and $T=10000$ K. Here $T$ is
excitation temperature of the gas describing the state population of via the Boltzmann law and assuming the local-thermal-equilibrium (LTE) and the temperature of the radiation field. The total cross section is obtained summing together the partial contributions. 
Table \ref{tab:temperatures} reports the individual temperatures. We assume that every excitation to an electronic state above the dissociation limits leads to a dissociative event. The same assumption is made by the Leiden database \citep{Leiden}.

At a given temperature $T$, the photodissociation cross section of a single electronic transition is calculated by averaging  100 individual cross sections produced by varying the radial grid size representing the range of the nuclear bond distances $r$. For each radial grid size $r_{\rm max}$, the Schr\"{o}dinger equation for the system is solved, where the unbound states are treated as a particle in box problem, on the basis of ``bound'' vibrational functions that vanish exactly at $r= r_{\rm max}$, as part of the Sinc DVR approach \citep{92CoMixx.method}. The resulting eigenvalues and eigenfunctions are then used to obtain temperature dependent cross sections  corresponding to the electronic, bound-bound and bound-free transitions.  The   bound-bound transitions are unaffected by a change in the radial grid, while the bound-free  transitions are sensitive to the increase grid size, as it increases the number of discrete ``particle-in-the-box'' states included in the Duo calculation. 
The single cross sections corresponding to different grid sizes are averaged and smoothed using a normalized Gaussian smoothing function.
The Schr\"{o}dinger equation is solved between $r_{\rm min}=0.5$ \AA\/ and $r_{\rm max}=3.0-3.1$~\AA with an increase of 0.001 \AA\/, 
 using a Sinc DVR basis set. In the Sinc DVR method, the interatomic distance is represented by a uniformly spaced grid points distributed  between $r_{\rm min}$ and $r_{\rm max}$; the Schr\"{o}dinger equation is transformed to an eigenmatrix problem. The real symmetric Hamiltonian matrix is then diagonalized for each $J$ (total angular momentum) and parity. Energies up to $hc$150~000 \invcm, $J_{\rm max}$ = 120  and vibrational levels up to $v= $200 are considered. For each grid size, a line list (energies and Einstein A coefficients) is produced with the thresholds of $10^{-30}$~cm molecule$^{-1}$ set for line intensities at $T=10~000$ K and of $10^{-30}$~D$^2$ for line strengths. Each line list is then used to calculate temperature dependent absorption coefficients (cm/molecule), binned into a wavenumber grid with the typical spacing of 4~\invcm. Each binned  spectrum for a given $T$  represents absorption transitions to a selection of unbound states corresponding to a given grid of $r_{\rm max}$. Different spectra are then averaged and smoothed with an appropriate Gaussian profile to produce the final absorption cross section (cm$^2/$molecule) for each temperature $T$. A Gaussian smoothing function is used with its width varying depending on the nature of the electronic transition. A half-width-of-half-maximum (HWHM) 
 of  12.5 \invcm\ was used for all bound-bound transitions ( C\,$^1\Pi \leftarrow$X\,$^1\Sigma^+$ for both molecules, D\,$^1\Pi\leftarrow$ X\,$^1\Sigma^+$, \, H$^1\Pi\leftarrow$X\,$^1\Sigma^+$, M\,$^1\Pi\leftarrow$X\,$^1\Sigma^+$, 4\,$^1\Sigma^+\leftarrow$X\,$^1\Sigma^+$ , and the 5\,$^1\Pi\leftarrow$X\,$^1\Sigma^+$ for HCl only) to reproduce the experimental cross sections by \citet{85NeSuLe.HF,02ChChBa.HCl} and \citet{06LiZhYu.HCl} for HCl and \citet{84HiWiBr.HF} for HF. 
The bound to dissociative  A\,$^1\Pi \leftarrow$X\,$^1\Sigma$ transitions of HCl and HF are smoothed using a Gaussian function with width a of 50 \invcm\  and 45 \invcm, respectively. A width of  125 \invcm  is used to describe the   B\,$^1\Sigma \leftarrow$X\,$^1\Sigma$ bound-bound transitions. The necessity of a large width for this transition derives from the fact that the B\,$^1\Sigma^+$ state is characterized by a small vibrational intervals and rotational constants \citep{71TiGixx.HCl}.

Calculations are performed on both hydrogen and deuterium isotopologues of  HCl and HF, and $^{35}$Cl and $^{37}$Cl isotopes for HCl. Isotopologue data are presented in Table \ref{tab:isotopes}. Each isotopologue cross section is presented for 100 \% abundance.

\begin{table}
\centering
\caption{Temperatures ($T$) at which the cross sections and rates presented in this work are calculated.}
\label{tab:temperatures}
\begin{tabular}{lcccr} 
\hline
  &  & $T$ (K) & & \\
\hline
0    	&100 	&200 	&300 	&400  \\
500    	&600 	&700 	&800 	&900  \\
1000	&1100 	&1200 	&1300 	&1400  \\
1500	&1600 	&1700 	&1800 	&1900  \\
2000	&2200 	&2400 	&2600 	&2800  \\
3000    &3200 	&3400 	&3600 	&4000  \\
4500    &5000 	&5500 	&10000 	&  \\
\hline
\end{tabular}
\end{table}

\begin{table}
\centering
\caption{HF and HCl isotopologues studied in this work.}
\label{tab:isotopes}
\begin{tabular}{lcccr} 
\hline
 Molecule & mass (Dalton) & nuclear spin \\
\hline
HF      	& 20.01	& $\frac{1}{2}$,$\frac{1}{2}$	\\
DF       	& 21.01	& $1$,$\frac{1}{2}$	            \\
\hline
H$^{35}$Cl	& 35.98	& $\frac{1}{2}$,$\frac{3}{2}$	\\
D$^{35}$Cl	& 36.98	& $1$,$\frac{3}{2}$	            \\ 
H$^{37}$Cl	& 37.97	& $\frac{1}{2}$,$\frac{3}{2}$	\\
D$^{37}$Cl  & 38.98	& $1$,$\frac{3}{2}$	            \\
\hline
\end{tabular}
\end{table}

\subsection{Radiation field and interstellar rates}

Photodissociation rates can be used as an alternative to the cross sections; they are used for modelling the abundance and temporal evolution of chemical species in the interstellar medium or in stellar and planetary atmospheres \citep{jt525,jt604}. 

The temperature-dependent photodissociation rate $k(T)$ of a molecule dissociated by a  field with a flux  $F(\lambda)$ between the wavelengths $\lambda_1$ and $\lambda_2$ is expressed as:
\begin{equation}\label{eq:radfield}
  k(T)=\int_{\lambda_1}^{\lambda_2} F(\lambda)\sigma(\lambda,T)d\lambda.
\end{equation}
Here and elsewhere $T$ refers to the temperature of the internal states of the molecule under consideration.

There are several standard fluxes used to produce appropriate rates, depending on which region of space is observed. The interstellar radiation field (ISRF) has been fitted by \citet{78Draine} to an analytical expression for wavelengths between 91.2 nm and 200 nm and was expressed as:
\begin{equation}
F(\lambda) = 3.2028\times 10^{13}\lambda^{-3} -5.1542\times10^{15} \lambda^{-4} + 2.0546\times10^{17} \lambda^{-5}  
\end{equation}
where $\lambda$ is the wavelength in nm, and it was later extended to 2000 nm by \citet{82DiBl} using the expression:
\begin{equation}
  F(\lambda) =3.67\times 10^4\lambda^{0.7}. 
\end{equation}
The stellar field is expressed in units of photons  s$^{-1}$ cm$^{-2}$ nm$^{-1}$.

The blackbody radiation field, $B(\lambda,T_{\rm rad})$, is used as an approximation of a generic stellar field at a temperature $T_{\rm rad}$, and expressed in units of photons s$^{-1}$ cm$^{-2}$ nm$^{-1}$. The general expression for this field is: 
\begin{equation}
    B(\lambda,T_{\rm rad}) =\frac{2 \times 10^4 }{ 4 \pi \lambda^4} \frac{1}{e^{\frac{hc}{\lambda k_{\rm B} T_{\rm rad}}} -1 }
\end{equation}
 $h=6.626 \times 10^{-34}$ J $\times$s is the Planck constant, $c$  is the speed of light, and  $k_{\rm B}=1.381 \times 10^{-23}$ J K$^{-1}$ is the Boltzmann constant. We  consider   several different temperatures, describing different types of stars, in a similar fashion to the work performed by \citet{Leiden}. The temperature $T_{\rm rad}=4000$ K is chosen for modelling the behaviour of T Tauri stars, stars that are in the early stages of formation and shine due to the gravitational energy of their collapse \citep{89ApMuxx,93Naxxxx}. The Herbig Ae stars, young A stars still embedded in gas dust envelope, are modelled using the blackbody temperature of $T_{\rm rad}=10~000$ K \citep{18ViOuBa}. The bright and short living B stars are modelled using a blackbody temperature of $T_{\rm rad}=20~000$ K \citep{81HaHexx}. 
 
Real fields present a series of absorption lines by species in the atmospheres. Here we consider real stellar fields for the Sun and  Proxima Centauri. The Solar field was taken from \citet{Leiden} and was compiled from the measurements of \citet{86WoPrRo} and \citet{01CuBrFe}; two Proxima Centauri fields are considered, the flux obtained using the PHOENIX model \citep{PHOENIX} and one
taken from the MUSCLES treasury survey \citep{MUSCLES_I,MUSCLES_II,MUSCLES_III,MUSCLES_IV}. The MUSCLES database gives the observational spectrum, reconstructed from different sources \citep{MUSCLES_IV} with various correction at low wavelengths. These two fields behave very differently both in shape of the fields themselves and, in the predicted photodissociation rates at essentially all temperatures. 
 We followed \citet{Leiden} and normalised all radiation field to agree with the integrated energy-intensity of the ISRF radiation field calculated by \cite{78Draine} between 91.2 and 200 nm ($I_D = 2.6 \times 10^{-10}$  W cm${^2}$).

\begin{figure*}
	\includegraphics[width=1.0\columnwidth]{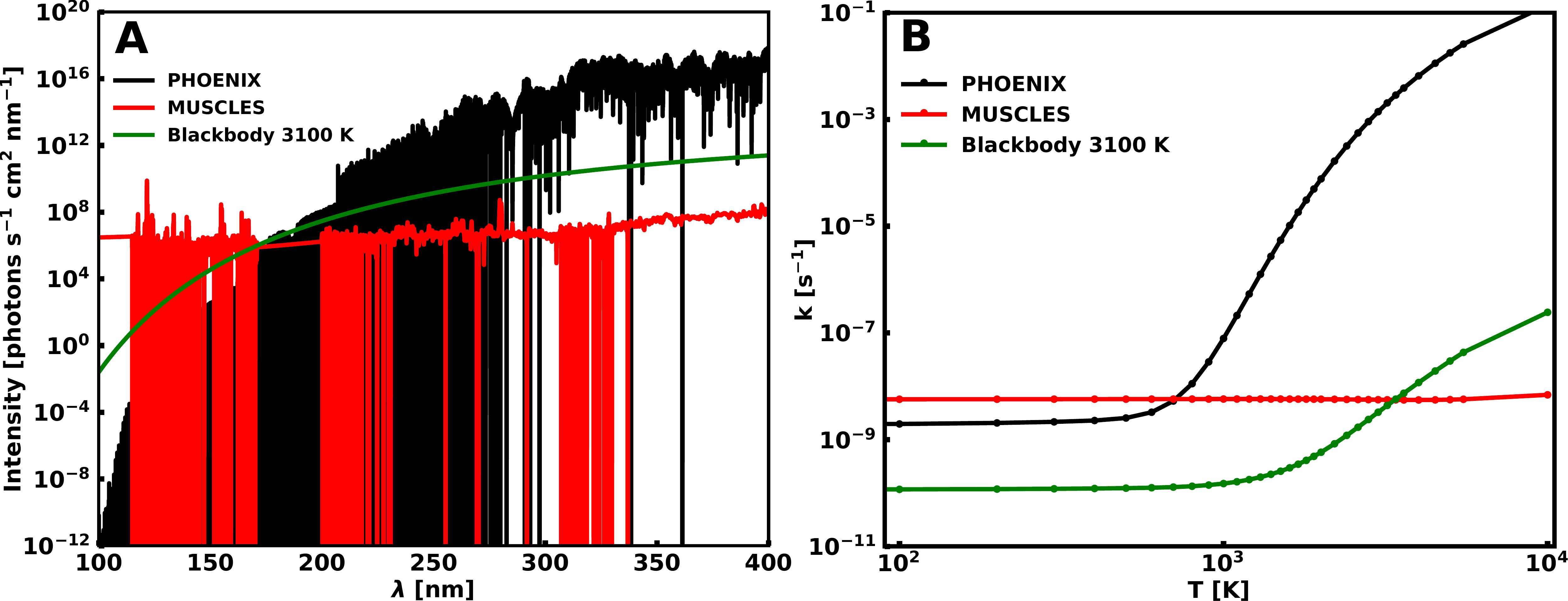}
    \caption{Panel A: different radiation fields models used for describing Proxima Centauri. The PHOENIX model \citep{PHOENIX} is  in black, MUSCLES model  \citep{MUSCLES_IV}   in red, and the blackbody in green.  Panel B: HCl photodissociation rates between 100 K and 10~000K using the different Proxima Centauri  stellar fields.}
    \label{fig:kHCl-proxima}
\end{figure*}

Figure \ref{fig:kHCl-proxima} compares  the PHOENIX and MUSCLES radiation fields for Proxima Centauri with a blackbody distribution as function of the wavelength, and the resulting HCl photodissociation rates. The MUSCLES and PHONEIX models differ from the blackbody model and each other; the MUSCLES field gives rates which are essentially temperature-independent while the PHOENIX field gives rates which increase   dramatically with temperature. This figure shows the necessity of using  reliable, high quality  UV-visible fluxes in any exoplanetary photochemical model.      

\section{HCl}

\begin{figure*}
	\includegraphics[width=\columnwidth]{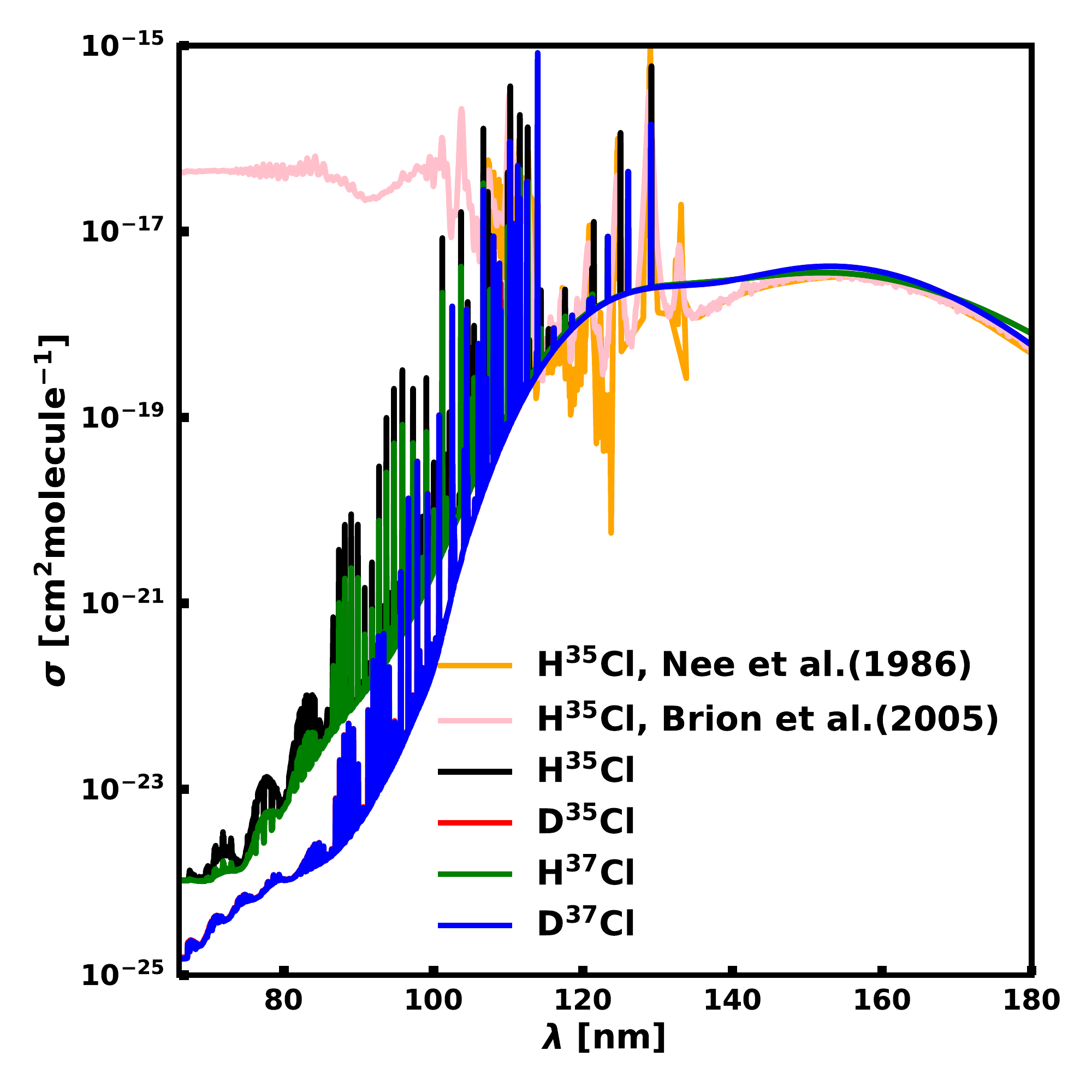}
    \caption{Total photodissociation cross section of the different isotopologues of HCl at $T=0$ K. 
    The cross section tails, for $\lambda < 100$ nm, show the difference between different hydrogen isotopologues: the deuterated species are consistently smaller. The D$^{35}$Cl spectrum is hidden by the D$^{37}$Cl spectrum.
    Experimental cross sections from \citet{86NeSuMa.HCl} and \citet{05BrDyCo.HCl} are also reported. The difference between the experiments and our model are largely explained by the use of different resolutions. In particular, bound states peak heights depend on the HWHM adopted: they can be several orders of magnitude higher for narrower averaging, however the integrated cross section is conserved.}
    \label{fig:HCl-isotopes}
\end{figure*}

The HCl photoabsorption spectrum  between 100 nm and 200 nm (see Fig.~\ref{fig:HCl-isotopes}) is dominated by three different regions: a continuum band centered at 153.64 nm given by the A\,$^1\Pi \leftarrow$X\,$^1\Sigma^+$ bound to continuum transitions, a second structured region, between 129 nm and 120 nm, with contributions from the B\,$^1\Sigma^+$ ionic bound state, the C\,$^1\Pi$, and D\,$^1\Pi$ bound states, and the last region, between 113 nm 104 nm, that has contributions from the H$^1\Sigma^+$, K\,$^1\Pi$, 5\,$^1\Pi$, M\,$^1\Pi$ and 4\,$^1\Sigma^+$ bound states. 
The experimental spectra from \citet{86NeSuMa.HCl} and \citet{05BrDyCo.HCl} are shown as well in Fig.~\ref{fig:HCl-isotopes}), showing an overall agreement with our model before smoothing is applied to the bound states. The difference between the experiments and our model are largely explained using different resolutions. In particular, bound states peak heights depend on the HWHM adopted: they can be several orders of magnitude higher for narrower averaging, however the integrated cross section is conserved.

The first region is dominated by the direct photodissociation from the A\,$^1\Pi$ repulsive electronic state. The A--X band  is associated with  the HCl$\rightarrow$H($^2$S)+Cl($^2$P) asymptote. The second region is mainly dominated by the vibrational progression of the C\,$^1\Pi$--X\,$^1\Sigma^+$ band  that overlaps with the transitions of the B\,$^1\Sigma^+$--X\,$^1\Sigma^+$ and D\,$^1\Pi$ --X\,$^1\Sigma^+$ photodissociation bands. The dominance of the C\,$^1\Pi \leftarrow$X\,$^1\Sigma^+$  band in this region is driven by its strong transition moment around the equilibrium region of the ground state potential. The B\,$^1\Sigma^+\leftarrow$X\,$^1\Sigma^+$ contribution shows continuum structure, due to the ionic nature of the dissociation limit of this state (HCl$\rightarrow$H$^+$+Cl$^-$), given by the contributions from the upper minimum  at 1.35 \AA\/. The third region is dominated by the K\,$^1\Pi\leftarrow$X\,$^1\Sigma^+$, M\,$^1\Pi\leftarrow$X\,$^1\Sigma^+$, 4\,$^1\Sigma^+\leftarrow$X\,$^1\Sigma^+$ bands, with the 5\,$^1\Pi\leftarrow$X\,$^1\Sigma^+$ contribution submerged, due to its small cross section. Given the complexity of the electronic bands in this region, it is not easy to associate single contributions to each of the vibrational progression. 

Table \ref{tab:hcl-h} compares our results with the experiments of \citet{86NeSuMa.HCl,02ChChBa.HCl} and \citet{06LiZhYu.HCl}, and  theoretical studies of \citet{82DiHeDa.HCl} and \citet{12EnSiCa.HCl} for H$^{35}$Cl. Table \ref{tab:hcl-d} presents a similar comparison with  experiment of \citet{02ChChBa.HCl} for D$^{35}$Cl. We assume a temperature of $T=0$ K for these comparisons and  use the oscillator strengths ($f$) instead of cross sections, in order to have a direct comparison with the previous studies. The conversion between cross sections and oscillator strength is made using the same formulation of \citet{02ChChBa.HCl}:
\begin{equation}
    f=1.13 \times 10^{12} \int^{\nu_2}_{\nu_1} \sigma(\nu)d\nu
\end{equation}
where $\nu_1$ and $\nu_2$ are the integration limits, in wavenumbers, for a given peak. 

The A\,$^1\Pi \leftarrow$X\,$^1\Sigma^+$ peak position  and $f$ agree within $\pm 0.19$ nm and $3.0\times10^{-3}$ respectively with the experimental values from \citet{02ChChBa.HCl}. There is a discrepancy between the experimental and calculated peak position values for D$^{35}$Cl, as our $\lambda_{\rm max}$ is blue-shifted by 1.74 nm.  The transitions involving the C\,$^1\Pi$ state are experimentally available for multiple vibrational states for the hydrogen isotopes. For both isotopologues,  our vibrational overtone positions of the excited state are more closely spaced than those observed experimentally. The oscillator strengths of the D\,$^1\Pi \leftarrow$X\,$^1\Sigma^+$, H$^1\Sigma \leftarrow$X\,$^1\Sigma^+$, K$^1\Pi \leftarrow$X\,$^1\Sigma^+$ , M$^1\Pi \leftarrow$X\,$^1\Sigma^+$, $4^1\Sigma \leftarrow$X\,$^1\Sigma^+$  bands agree with the results of \citet{12EnSiCa.HCl}.  However, the oscillator strength of the $5^1\Pi \leftarrow$X\,$^1\Sigma^+$ band is two orders of magnitude lower than the values of \citet{12EnSiCa.HCl}.

\begin{landscape}
\begin{table*}
\centering
\caption{Comparison between electronic transitions of HCl presented by \citet{82DiHeDa.HCl}, \citet{86NeSuMa.HCl}, \citet{02ChChBa.HCl}, \citet{06LiZhYu.HCl}, \citet{12EnSiCa.HCl} and our values. For each electronic state the vibronic progression, the peak position the oscillator strength, $f$, are given.}
\label{tab:hcl-h}
\resizebox{1.4\textwidth}{!}{
\begin{tabular}{lccccccccccccr} 
\hline
State &	Transition & $\lambda_{\rm van Dishoeck}$ (nm) & $\lambda_{\rm Nee}$ (nm) & $\lambda_{\rm Cheng}$ (nm) & $\lambda_{\rm Li}$ (nm) & $\lambda_{\rm Engin}$ (nm) & $\lambda_{\rm}$ (nm) &$f_{\rm van Dishoeck}$(unitless)  &$f_{\rm Nee}$(unitless) &$f_{\rm Cheng}$(unitless)  &$f_{\rm Li}$(unitless) &$f_{\rm Engin}$(unitless)  &	$f$ (unitless) \\
\hline
A\,$^1\Pi$    	&broad          	&-     &-     &153.83  &-     &156.94   &153.64   &-                   &-                   &5.1$\times 10^{-2}$ &4.2$\times 10^{-2}$   &7.06$\times 10^{-2}$   &5.4$\times 10^{-2}$\\ 
\hline
C\,$^1\Pi$ 	    &$0 \rightarrow 0$ 	&128.88 &129.02 &129.15 &129.02 &131.20 &129.16 &1.5$\times 10^{-1}$ &1.3$\times 10^{-1}$ &1.3$\times 10^{-1}$ 	&1.5$\times 10^{-1}$  &1.4$\times 10^{-1}$  &1.5$\times 10^{-1}$ \\ 
            	&$0 \rightarrow 1$ 	&124.61 &124.36 &124.73 &124.73 &127.03 &125.03 &2.4$\times 10^{-2}$ &1.7$\times 10^{-2}$ &1.7$\times 10^{-2}$ 	&2.0$\times 10^{-2}$  &3.0$\times 10^{-2}$ 	&2.7$\times 10^{-2}$  \\ 
            	&$0 \rightarrow 2$ 	&120.84 &120.72 &120.84 &120.84 &123.37 &121.45 &2.4$\times 10^{-3}$ &2.6$\times 10^{-3}$ &2.6$\times 10^{-3}$  &4.1$\times 10^{-1}$  &5.6$\times 10^{-3}$  &2.7$\times 10^{-3}$ \\ 
            	&$0 \rightarrow 3$ 	&117.41 &-      &-      &-      &120.14 &118.27 &6.0$\times 10^{-5}$ &-                   &-                    &-                    &8.0$\times 10^{-4}$ 	&1.6$\times 10^{-5}$ \\ 
\hline
B\,$^1\Sigma^+$  	&broad              &-      &-      &-      &-      &130.92 &128.68 &-                   &-                   &-                     &- 	              &0.0                  &3.1$\times 10^{-2}$ \\
\hline
D\,$^1\Pi$ 	    &$0 \rightarrow 0$  &-      &-      &-      &-      &124.61 &121.24 &-                   &-                   &-                     &-                   &9.0$\times 10^{-4}$  &6.3$\times 10^{-4}$  \\ 
\hline
H$^1\Sigma^+$ 	&$0 \rightarrow 0$ 	&-      &112.61 &-      &112.82 &114.69 &112.61 &-                   &1.0$\times 10^{-2}$ &-                    &1.1$\times 10^{-2}$  &3.2$\times 10^{-2}$ 	&3.2$\times 10^{-2}$ \\ 
     	        &$0 \rightarrow 1$ 	&-      &-      &-      &-      &112.41 &109.94 &-                   &-                   &-                     &-                   &7.8$\times 10^{-3}$ 	&1.0$\times 10^{-2}$  \\ 
    	        &$0 \rightarrow 2$ 	&-      &-      &-      &-      &110.21 &107.59 &-                   &-                   &-                     &-                   &3.3$\times 10^{-4}$ 	&2.0$\times 10^{-4}$ \\ 
\hline
K\,$^1\Pi$ 	    &$0 \rightarrow 0$ 	&-     &111.40  &-      &111.60 &114.27 &111.56 &-                   &2.1$\times 10^{-2}$ &-                     &-                   &2.7$\times 10^{-2}$  &4.3$\times 10^{-2}$  \\ 
 	         	&$0 \rightarrow 1$ 	&-     &-       &-      &-      &111.00 &108.07* &-                   &-                   &-                     &-                   &2.3$\times 10^{-3}$ 	&1.4$\times 10^{-3}$ \\ 
\hline
5\,$^1\Pi$ 	    &$0 \rightarrow 0$  &-     &-       &-	    &-      &113.85 &111.00  &-                   &-                   &-                     &-                   &5.0$\times 10^{-5}$ &8.0$\times 10^{-7}$  \\ 
\hline
M\,$^1\Pi$ 	    &$0 \rightarrow 0$  &-     &110.21  &-	    &110.31 &113.23 &110.27 &-                   &3.2$\times 10^{-2}$ &-                     &1.3$\times 10^{-1}$ &8.5$\times 10^{-2}$ &8.7$\times 10^{-2}$ \\ 
 	         	&$0 \rightarrow 1$ 	&-     &-       &-      &-      &110.11 &107.28 &-                   &-                   &-                     &-                   &6.4$\times 10^{-3}$ &6.3$\times 10^{-3}$  \\ 
	 	        &$0 \rightarrow 2$ 	&-     &-       &-      &-      &107.35 &104.58 &-                   &-                   &-                     &-                   &5.0$\times 10^{-4}$ &4.1$\times 10^{-4}$ \\ 
\hline
4\,$^1\Sigma^+$ 	&$0 \rightarrow 0$ 	&-     &-       &-      &-      &110.60 &106.68  &-                   &-                   &-                     &-                   &3.2$\times 10^{-2}$ &3.0$\times 10^{-2}$  \\ 
\hline
\end{tabular}
}
\end{table*}
\end{landscape}
\begin{table*}
\centering
\caption{Comparison between electronic transitions presented by \citet{02ChChBa.HCl}, and our values for DCl. For each electronic state  the vibronic progression, the peak position, and  the oscillator strength, $f$, are given.}
\label{tab:hcl-d}
\begin{tabular}{lccccccr} 
\hline
State &	Transition & $\lambda_{\rm Cheng}$ (nm) & $\lambda_{\rm}$ (nm) &$f_{\rm Cheng}$(unitless) &	$f$ (unitless) \\
\hline
A\,$^1\Pi$    	&broad            	&155.76  &154.02  &4.9$\times 10^{-2}$ 	&5.4$\times 10^{-2}$\\ 
\hline
C\,$^1\Pi$ 	    &$0 \rightarrow 0$ 	&129.15 &129.11  &1.3$\times 10^{-1}$ 	&1.3$\times 10^{-1}$ \\ 
            	&$0 \rightarrow 1$ 	&125.87 &126.06  &1.7$\times 10^{-2}$ 	&3.9$\times 10^{-2}$ \\ 
            	&$0 \rightarrow 2$ 	&122.88 &123.37  &3.3$\times 10^{-3}$ 	&6.0$\times 10^{-3}$ \\ 
            	&$0 \rightarrow 3$ 	&120.26 &120.84  &7.0$\times 10^{-4}$ 	&3.0$\times 10^{-4}$ \\ 
\hline
D\,$^1\Pi$ 	    &$0 \rightarrow 0$  &121.20 &121.20  &6.0$\times 10^{-4}$ 	&4.9$\times 10^{-4}$  \\ 
\hline
\end{tabular}
\end{table*}

The effects of isotopic substitution on the photoabsorption spectrum of HCl are shown in Figure \ref{fig:HCl-isotopes}. The major isotopic effect is present in the short wavelength tail of the photoabsorption spectrum. The cross sections for hydrogenated species are systematically higher than deuterated for $\lambda \leq 110$ nm, and the spacing of the vibrational progression of hydrogenated species is higher than that encountered for deuterated species.  

\begin{figure*}
	\includegraphics[width=1.0\columnwidth]{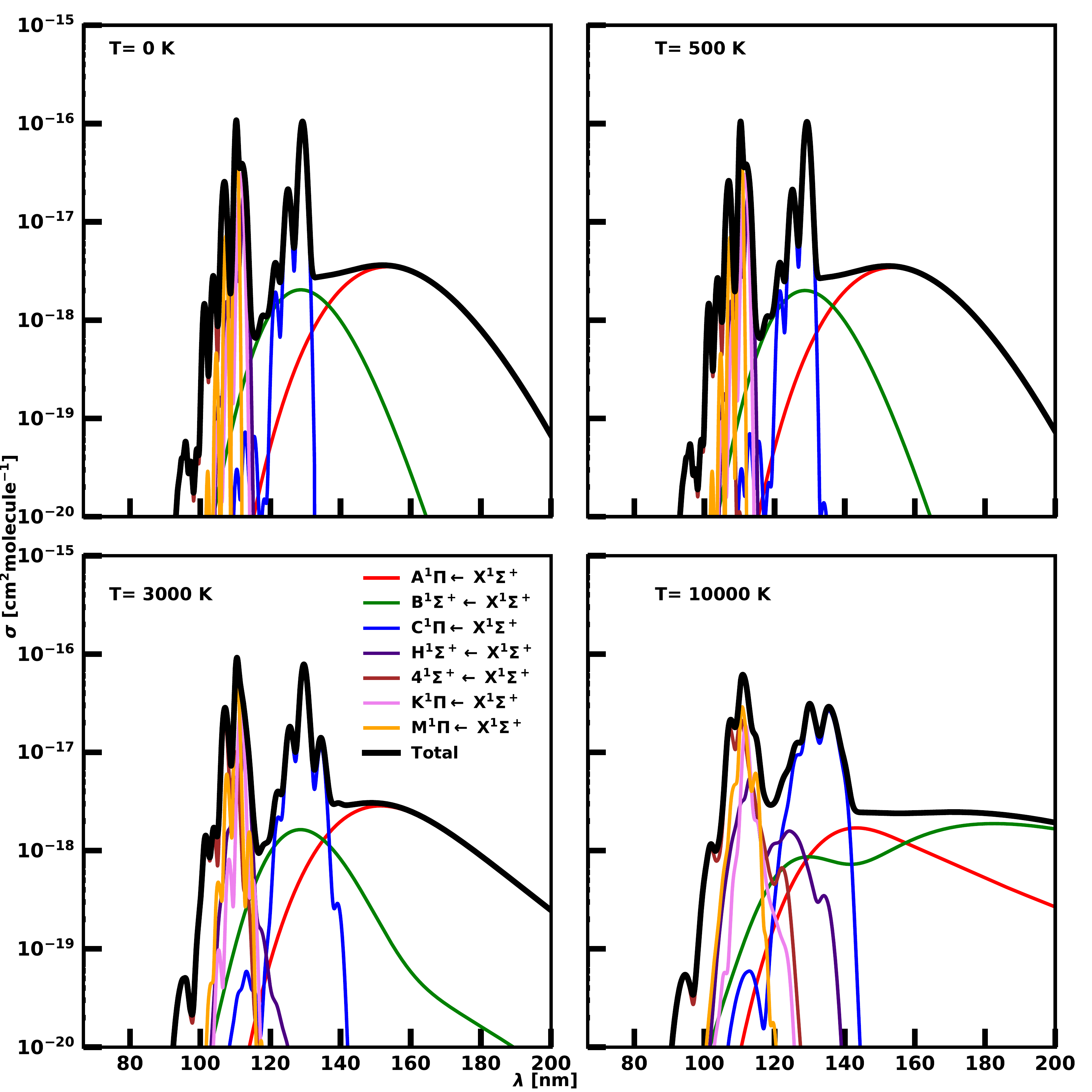}
    \caption{Total and partial photodissociation cross sections of H$^{35}$Cl at 0 K, 500 K, 3000 K. and 10000 K. The D\,$^1\Pi$ and the 5\,$^1\Pi$ states are excluded, due to their contributions being below $10^{-18}$ \xsec. The total spectrum is composed of a broad band given by the A\,$^1\Pi \leftarrow$X\,$^1\Sigma^+$ at 153.64 nm, a discrete progression derived by the C\,$^1\Pi \leftarrow$X\,$^1\Sigma^+$ at 129.16 nm and another set of discrete transitions starting at 112.61 nm, given by the contribution of the other excited states.}
    \label{fig:1H35Cl}
\end{figure*}

Figure \ref{fig:1H35Cl} presents the temperature dependence of the partial and total cross sections, by showing  cross sections at $T=0$ K, $T=500$ K, $T=3000$ K, $T=10~000$ K. The  D\,$^1\Pi\leftarrow$X\,$^1\Sigma^+$ and the 5\,$^1\Pi\leftarrow$X\,$^1\Sigma^+$ bands are excluded from the figure for sake of clarity, because their contributions are below $10^{-18}$ \xsec. There is no major difference between the cross sections at $T=0$ K and $T=500$ K, while major changes are observed above $T=3000$ K. There is an increase of the cross section tails at long wavelengths, as more excited vibrational states are accessible at high temperatures. At $T=10~000$ K,  the  $B^1\Sigma^+\leftarrow$X\,$^1\Sigma^+$ band shows two peaks. They correspond to two different minima of the $B^1\Sigma^+$ electronic state potential, as the minimum at $r_{\rm HCl}=2.44$ \AA\/ becomes accessible at high temperatures. Increasing the temperature results in a  degradation of the vibrational progressions and to a general redshift, of the order of $10-15$ nm for the C\,$^1\Pi\leftarrow$X\,$^1\Sigma^+$ transitions. The same group of transitions shows a change of the peak intensities in the vibrational progression, as the C\,$^1\Pi(\nu=1) \leftarrow$ X\,$^1\Sigma^+(\nu=0)$ transition becomes stronger than the C\,$^1\Pi(\nu=0)\leftarrow$X\,$^1\Sigma^+(\nu=0)$  at $T=3000$ K. The vibrational progression at 114.69 nm shows a general redshift for $T \leq 3000K$, with the effect changing depending on the electronic state under examination: the K\,$^1\Pi \leftarrow$X\,$^1\Sigma^+$ and M\,$^1\Pi \leftarrow$X\,$^1\Sigma^+$ bands show a shift up to 5 nm, the 4\,$^1\Sigma^+ \leftarrow$X\,$^1\Sigma^+$ band increases up to 20 nm, while the  H$^1\Sigma^+ \leftarrow$X\,$^1\Sigma^+$ band shifts for values above 25 nm. 

\begin{figure*}
	\includegraphics[width=1.0\columnwidth]{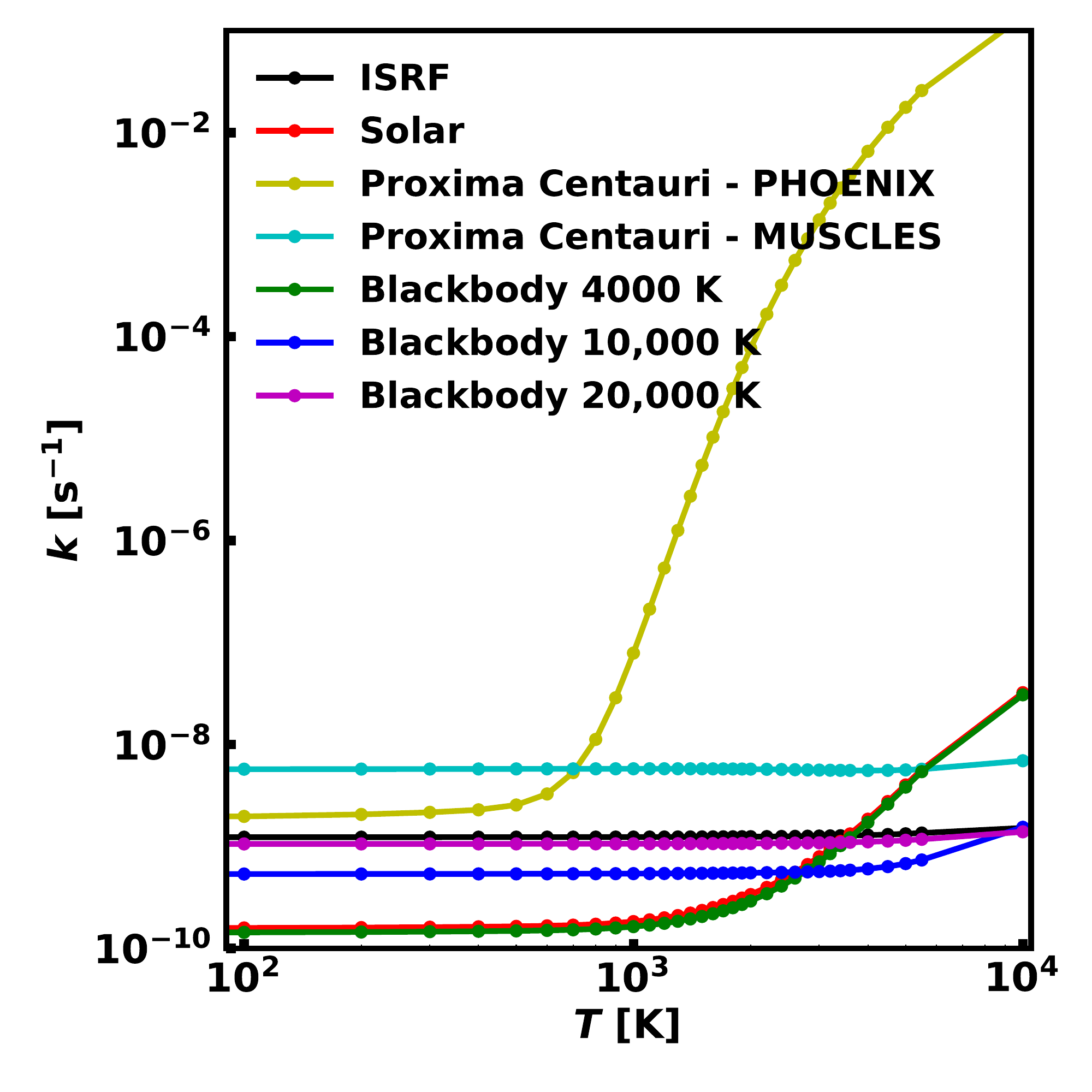}
    \caption{Photodissociation rates of H$^{35}$Cl between 100 K and 10~000K for different stellar fields. The ISRF is in black, the Solar field in red, the Proxima Centauri field with the PHOENIX model  \citep{PHOENIX} in yellow, the Proxima Centauri field with the MUSCLES model \citep{MUSCLES_IV} in cyan; the blackbody temperatures are in green (4000 K), in blue (10~000 K), and in violet (20~000 K). The temperature dependence of the photodissociation rates depends strongly on the stellar field, the higher the stellar field temperature, the lower is the increase in  photodissociation rate as function of molecular temperature. The photodissociation rates for the ISRF and the blackbody at 20~000 K do not show appreciable temperature dependence, while the Solar and the blackbody at 4000 K increase by a factor of 213 within the temperature interval considered. The rates obtained using the Proxima Centauri field with the PHOENIX model show a dramatic temperature effect, with a change in rates of the order of 10$^7$.}
    \label{fig:kHCl-1-35}
\end{figure*}

Figure \ref{fig:kHCl-1-35} shows how  the photodissociation rates change as function of temperature for H$^{35}$Cl. The stellar field influences the temperature dependence of the photodissociation rate. Fields due to high temperature stars tend to flatten the temperature effects on the rates: for ISRF and 20000 K they are constant within 10\% for all practical purposes. Considering the blackbody radiation at 4000 K, the photodissociation rate increases $213$ times from 100 K to 10~000 K. For the same temperature interval, the radiation field generated by a blackbody at 20~000 K shows an increase by a factor of only $1.32$. The most dramatic temperature effect is observed for the PHOENIX model stellar field of Proxima Centuari: the curve shows an exponential behaviour for temperatures above 1000 K; this effect leads to an increase by $7.36 \times 10^{7}$ times from 100~K to 10~000 K. The MUSCLES field for Proxima Centauri shows  completely different behaviour to the PHOENIX model, with essentially no  variation with  temperature.   

The effect of isotopic substitution is illustrated in Table \ref{tab:rateshcl}. H$^{37}$Cl and D$^{37}$Cl show similar temperature behaviour to H$^{35}$Cl in Figure \ref{fig:kHCl-1-35} for all stellar fields. The rates of D$^{35}$Cl using the 4000 K blackbody field are anomalous compared  to the others, as it shows a rapid increase with temperature giving a rate one order of magnitude higher at 10~000 K.

\citet{82DiHeDa.HCl} estimated  photodissociation rates  in the ISRF stellar field for HCl. Their model, which  included  5 singlet states and 5 triplet states, gave a total photodissociation rate of $9.81\times 10^{-10}$ s$^{-1}$, lower than our value of $1.23\times 10^{-9}$ s$^{-1}$. This difference is due to the exclusion in their models of excited states above 113 nm. They reported the single state contribution from the A\,$^1\Pi \leftarrow$X\,$^1\Sigma$, and the C\,$^1\Pi \leftarrow$X\,$^1\Sigma$ transitions.  Their values, $2.1 \times 10^{-10}$ s$^{-1}$ and  $5.3 \times 10^{-10}$ s$^{-1}$ respectively,  agree within the $10\%$ with our values, $2.3 \times 10^{-10}$ s$^{-1}$ and  $5.2 \times 10^{-10}$ s$^{-1}$, respectively. 
The total photodissociation rates at $T=0$ K are in good agreement with those reported by \citet{Leiden}. Our rates for the ISRF, Solar, and the blackbody at 4000 K, 10,000 K and 20,000 K are  $1.23 \times 10^{-9}$ s$^{-1}$, $1.57 \times 10^{-10}$ s$^{-1}$, $1.43 \times 10^{-10}$ s$^{-1}$, $5.36 \times 10^{-10}$ s$^{-1}$ and $1.06 \times 10^{-9}$ s$^{-1}$, versus $1.73 \times 10^{-9}$ s$^{-1}$, $1.08 \times 10^{-10}$ s$^{-1}$, $9.35 \times 10^{-11}$ s$^{-1}$, $5.06 \times 10^{-10}$ s$^{-1}$ and $1.47 \times 10^{-9}$ s$^{-1}$ reported in the database.

\begin{table*}
\centering
\caption{Photodissociation rates for all HCl  isotopologues at blackbody temperatures of   100 K, 1500 K, 3000 K, 5000 K and 10~000 K. }
\label{tab:rateshcl}
\begin{tabular}{lcccccr} 
\hline
Blackbody field &	Isotopologue & $k$(100 K) (s$^{-1}$) & $k$(1500 K) (s$^{-1}$) & $k$(3000 K) (s$^{-1}$) & $k$(5000 K) (s$^{-1}$) & $k$(10000 K) (s$^{-1}$) \\
\hline
 4000 K   & H$^{35}$Cl&$1.44\times 10^{-10}$&$2.06\times 10^{-10}$&$7.07\times 10^{-10}$&$3.81\times 10^{-9}$&$3.07\times 10^{-8}$  \\
                    & D$^{35}$Cl&$1.14\times 10^{-10}$&$3.34\times 10^{-9}$ &$2.32\times 10^{-8}$ &$6.51\times 10^{-8}$&$2.04\times 10^{-7}$  \\
                    & H$^{37}$Cl&$1.43\times 10^{-10}$&$2.06\times 10^{-10}$&$7.07\times 10^{-10}$&$3.81\times 10^{-9}$&$3.07\times 10^{-8}$  \\
                    & D$^{37}$Cl&$1.08\times 10^{-10}$&$1.76\times 10^{-10}$&$6.70\times 10^{-10}$&$3.74\times 10^{-9}$&$3.00\times 10^{-8}$  \\
\hline
10~000 K & H$^{35}$Cl&$5.37\times 10^{-10}$&$5.46\times 10^{-10}$&$5.67\times 10^{-10}$&$6.79\times 10^{-10}$&$1.54\times 10^{-9}$  \\
                    & D$^{35}$Cl&$5.46\times 10^{-10}$&$5.83\times 10^{-10}$&$7.03\times 10^{-10}$&$9.64\times 10^{-10}$&$2.04\times 10^{-9}$   \\
                    & H$^{37}$Cl&$5.35\times 10^{-10}$&$5.45\times 10^{-10}$&$5.67\times 10^{-10}$&$6.79\times 10^{-10}$&$1.54\times 10^{-9}$  \\
                    & D$^{37}$Cl&$5.37\times 10^{-10}$&$5.47\times 10^{-10}$&$5.69\times 10^{-10}$&$6.79\times 10^{-10}$&$1.53\times 10^{-9}$  \\
\hline
20~000 K & H$^{35}$Cl&$1.05\times 10^{-9}$&$1.07\times 10^{-9}$ & $1.08\times 10^{-9}$&$1.15\times 10^{-9}$&$1.39\times 10^{-9}$  \\
                    & D$^{35}$Cl&$1.12\times 10^{-9}$&$1.13\times 10^{-9}$ & $1.16\times 10^{-9}$&$1.25\times 10^{-9}$&$1.50\times 10^{-9}$  \\
                    & H$^{37}$Cl&$1.06\times 10^{-9}$&$1.07\times 10^{-9}$ & $1.08\times 10^{-9}$&$1.15\times 10^{-9}$&$1.39\times 10^{-9}$  \\
                    & D$^{37}$Cl&$1.07\times 10^{-9}$&$1.07\times 10^{-9}$ & $1.09\times 10^{-9}$&$1.15\times 10^{-9}$ & $1.40\times 10^{-9}$  \\
\hline
\end{tabular}
\end{table*}

\section{HF}

The photodissocation spectrum of HF is characterized by two regions: the continuum band centered at 120.09 nm given by the A\,$^1\Pi \leftarrow$X\,$^1\Sigma^+$ transitions, the second structured region between 95 nm and 80 nm, with  the main contribution from the C\,$^1\Pi$ state, and the  B\,$^1\Sigma^+$--X\,$^1\Sigma^+$ band, that is submerged below the C--X band. 

 Experiments reporting the HF photoabsorption are due to \citet{84HiWiBr.HF}, who describes different electronic transitions, and \citet{85NeSuLe.HF} who reports cross section for the A\,$^1\Pi \leftarrow$X\,$^1\Sigma^+$ only. The MPI-Mainz database also cites the experimental data from \citet{81CaTsBr.HF} and host their photoabsorption cross sections of HF, the agreement with their data is less good, given lower experimental resolution. Table \ref{tab:hf} compares our peak positions and oscillator strengths of HF  with  experimental data  from \citet{84HiWiBr.HF}. Our calculations agree with the  A\,$^1\Pi \leftarrow$X\,$^1\Sigma^+$ band  reported by \citet{85NeSuLe.HF}  with a peak position at $\lambda=121.70$ nm, cross section of $3.3 \times 10^{-18}$ \xsec and $f=4.6\times 10^{-2}$.  Nee {\it et al.}'s measurements give a cross sections a factor of 0.53 lower than  the experiments of \citet{84HiWiBr.HF},  but which agrees with our calculations  and with the theoretical results of \citet{00BrBa.HF}.  Our peak positions and oscillation strengths for the C\,$^1\Pi \leftarrow$X\,$^1\Sigma^+$ transitions are in a good agreement with \citet{84HiWiBr.HF}, while the oscillator strengths for the  A\,$^1\Pi \leftarrow$X\,$^1\Sigma^+$ and the B\,$^1\Sigma^+ \leftarrow$X\,$^1\Sigma^+$ differ by a factor of 0.50 and 0.56 respectively, supporting the experiments of \citet{85NeSuLe.HF}. Considering the good agreement between our methodology and experiments for HCl, we suppose that the experimental data from \citet{84HiWiBr.HF} could have some problem in assigning transitions of different nature (continuum \textit{versus} discrete). 

\begin{table*}
\centering
\caption{Comparison between electronic transitions presented by \citet{84HiWiBr.HF}, and our values. for the deuterated species. For each electronic state we are reporting the vibronic progression, the peak position, and  the oscillator strength, $f$.}
\label{tab:hf}
\begin{tabular}{lccccccr} 
\hline
State &	Transition & $\lambda_{\rm Hitchcock}$ (nm) & $\lambda_{\rm}$ (nm) &$f_{\rm Hitchcock}$(unitless) &	$f$ (unitless) \\
\hline
A\,$^1\Pi$    	&broad           	&119.79 &119.78  &9.8$\times 10^{-2}$ 	&4.9$\times 10^{-2}$\\ 
B\,$^1\Sigma^+$  &broad            	&93.93   &93.93  &2.5$\times 10^{-2}$ 	&1.4$\times 10^{-2}$ \\ 
C\,$^1\Pi$ 	   &$0 \rightarrow 0$ 	&95.15   &95.17  &5.6$\times 10^{-2}$ 	&5.7$\times 10^{-2}$ \\ 
\hline
\end{tabular}
\end{table*}

\begin{figure*}
	\includegraphics[width=1.0\columnwidth]{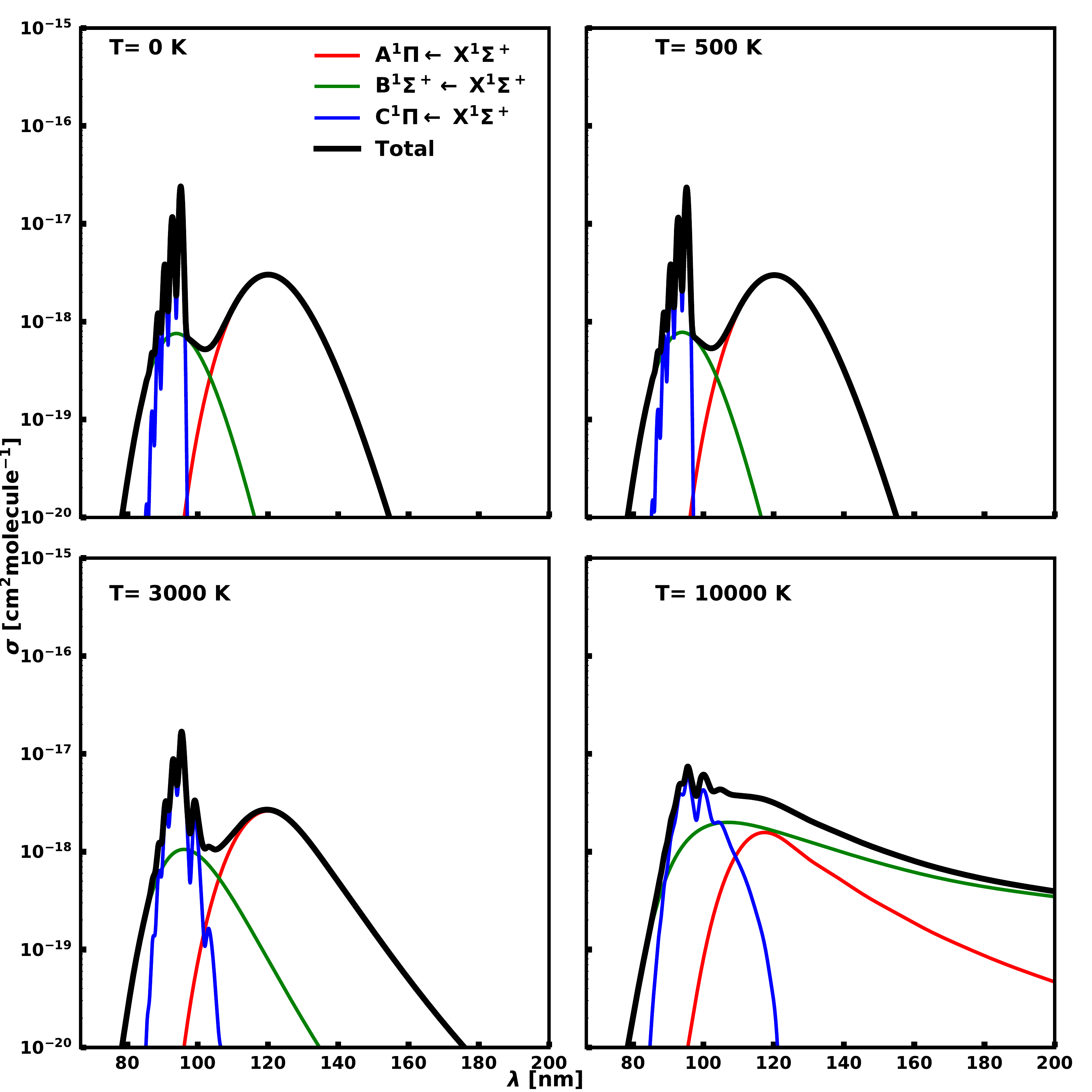}
    \caption{Total and partial photodissociation cross sections of HF at 0 K, 500 K, 3000 K. and 10000 K. The total spectrum is composed of a broad  band given by A\,$^1\Pi \leftarrow$X\,$^1\Sigma^+$  at 120.09 nm, and a discrete progression of C\,$^1\Pi \leftarrow$X\,$^1\Sigma^+$ at 95.17 nm.}
    \label{fig:1H19F}
\end{figure*}

The temperature effect on the HF cross sections is illustrated  in Figure \ref{fig:1H19F}. As in the case of HCl, there are no major differences between spectra between 0 and 500 K. We observe a change in the C\,$^1\Pi \leftarrow$X\,$^1\Sigma^+$ vibrational progression, that shifts to energy values above 100 nm, and the increase of the energy tail for high wavelengths, mainly from the A\,$^1\Pi \leftarrow$X\,$^1\Sigma^+$ band. 

\begin{figure*}
	\includegraphics[width=\columnwidth]{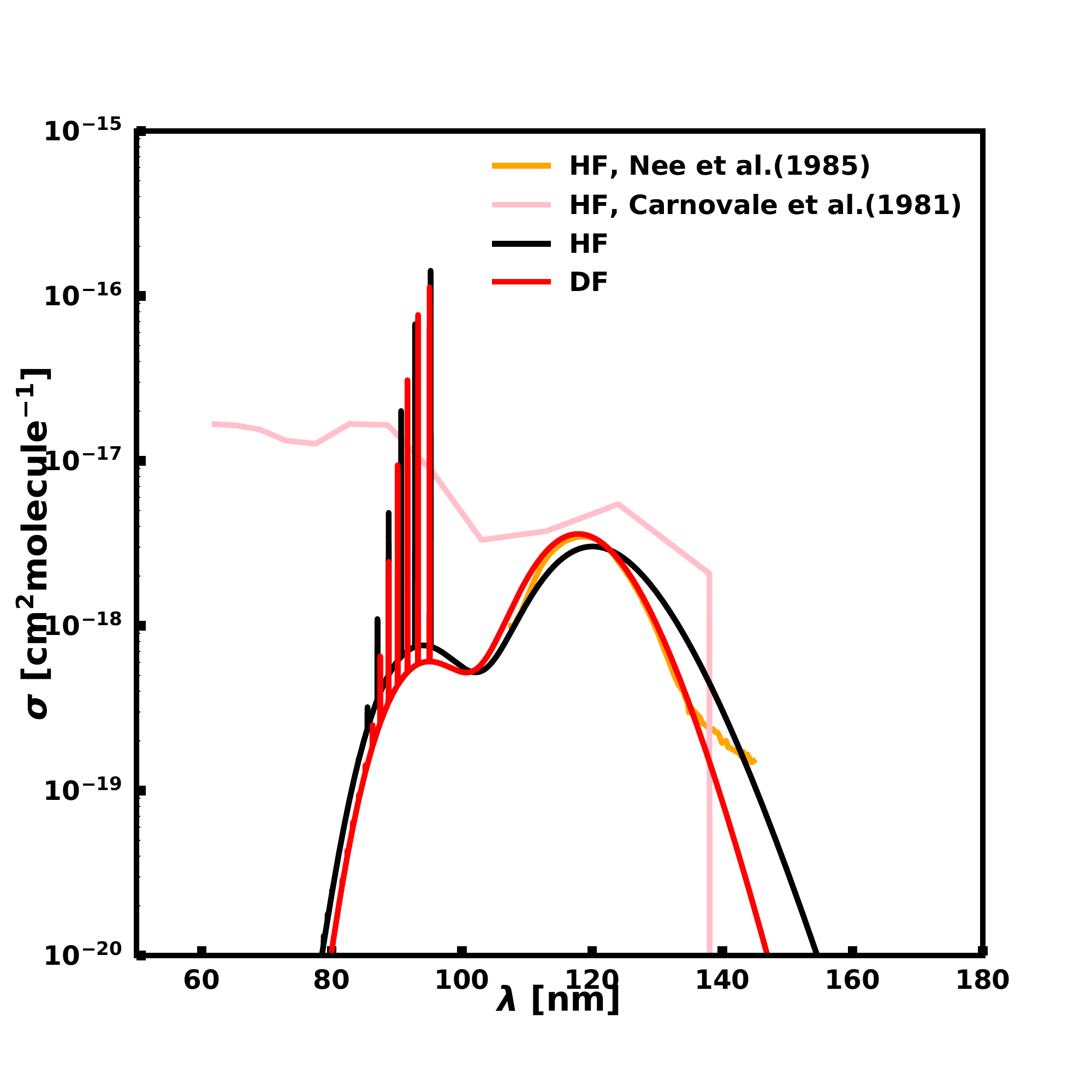}
    \caption{Total photodissociation cross section of HF and DF at $T=0$ K. The A\,$^1\Pi \leftarrow$X\,$^1\Sigma^+$ band shows a shift of 2.20 nm between the hydrogenated and the deuterated species. The spacing between the C\,$^1\Pi \leftarrow$X\,$^1\Sigma^+$ vibrational progression is smaller than that observed in  HF. These spectra are compared with the experimental results from \citet{81CaTsBr.HF} and \citet{85NeSuLe.HF}; in both cases the scarcity of experimental data is evident: in case of \citet{81CaTsBr.HF} the sampling is too coarse to discern the vibrational progressions from the bound states, while \citet{85NeSuLe.HF} reports the contribution of the A\,$^1\Pi \leftarrow$X\,$^1\Sigma^+$ band. Bound states peak heights depend on the Gaussian smoothing function adopted: they can be several orders of magnitude higher with narrower averaging, however the integrated cross section is conserved. }
    \label{fig:HF-isotopes}
\end{figure*}

The computational work of \citet{00BrBa.HF} is the only source of information available for the  A\,$^1\Pi \leftarrow$X\,$^1\Sigma^+$ band of DF. They reported that the photodissociation cross section of DF peaked at higher wavelengths with a higher peak and narrower profile. Our data shows a similar behaviour, with a difference of 2.20 nm between the cross sections of HF  and DF. The spacing between the C\,$^1\Pi \leftarrow$X\,$^1\Sigma^+$ vibrational progression is smaller than in HF, due to the increase of mass of the deuterated species. 
Figure \ref{fig:HF-isotopes} shows the the cross sections of the two isotopes together with the experimental data from \citet{81CaTsBr.HF} and \citet{85NeSuLe.HF}.

\begin{figure*}
	\includegraphics[width=1.0\columnwidth]{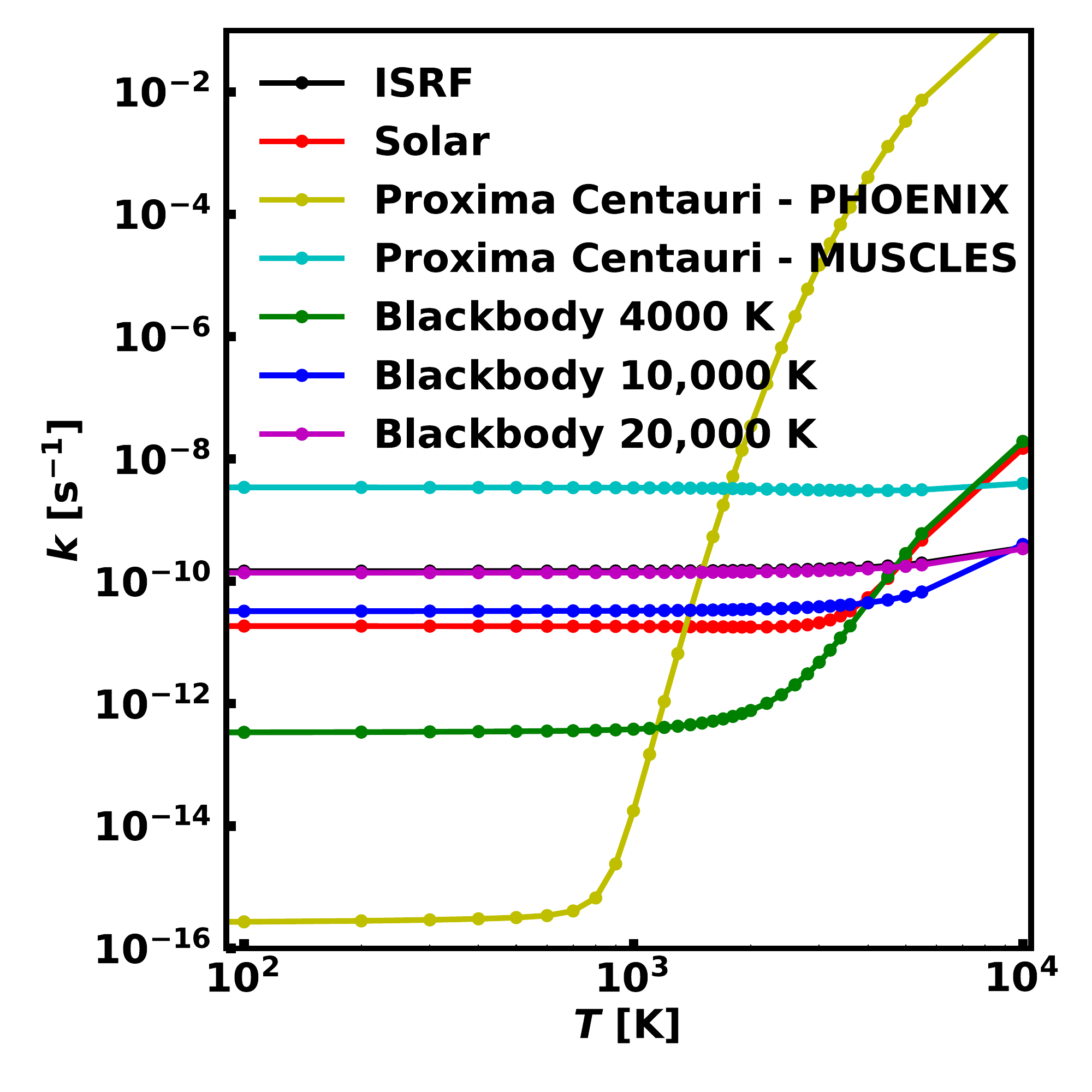}
    \caption{Photodissociation rates between 100 K and 10~000K for different stellar fields of HF. The ISRF is in black, the Solar field in red, the Proxima Centuary with the PHOENIX model  \citep{PHOENIX}  in yellow, the Proxima Centuary with the MUSCLES model \citep{MUSCLES_IV} in cyan; the blackbody temperatures are in green (4000 K), in blue (10~000 K), and in violet (20~000 K). The temperature dependence of the photodissociation rates depends strongly on the stellar field, the higher is the stellar field temperature, the lower is the increase of the photodissociation rate as function of temperature. The photodissociation rates for the ISRF and the blackbody at 20~000 K and the MUSCLES model  is essentially independent of temperature, while the Solar and the blackbody at 4000 K increase of a factor of $\sim$ 250 within the temperature interval. The ISRF and the blackbody at 20~000 K curves overlap through the entire temperature interval. The Proxima Centauri field with the PHOENIX model shows the strongest temperature dependence, with an increase of  $9 \times 10^{13}$ times within the temperature range. }
    \label{fig:kHF-1-19}
\end{figure*}

HF photodissociation rates for different radiation fields are plotted in Figure \ref{fig:kHF-1-19}. The temperature dependence of the blackbody field at $20~000$ K and the ISRF is similar to those observed for HCl: they have a similar temperature dependence, and increase  by a factor of 2.66 as $T$ increases from  $100$ K to $10~000$ K. More significant differences between HCl and HF are encountered for the blackbody radiation field at $4000$ K and the Proxima Centauri fields. For temperatures below $2000$ K the $4000$ K field is two orders of magnitude smaller than the solar field, in contrast to the case of HCl, where the fields have the same trend. There is a sudden increase in the rates from $2000$ K, with the Solar and the $4000$ K blackbody field showing the same trend after $4000$ K. Rates computed using the Proxima Centauri field with the PHOENIX model show the greatest increase in the temperature range: $9 \times 10^{13}$ times, from $100$ K to $10~000$ K.

Table \ref{tab:rateshf} shows the isotopic dependence of the photodissociation rate of HF and DF. For both isotopes, the  $10~000$ K and $20~000$ K blackbody fields have the same temperature dependence photodissociation rates. In case of the blackbody field at 4000 K, we observe that the deuterated isotopologue shows rates one order of magnitude lower than for HF, for temperatures below 3000 K, but rapidly increases for higher temperature: the ratio between $k_{\rm DF}(10~000)$ and $k_{\rm HF}(10~000)$ is 0.61. Also, for HF,  total photodissociation rates at $T=0$ K agree with the ones reported by \citet{Leiden}. Our rates for the ISRF, Solar, and the blackbody at 4000 K, 10,000 K and 20,000 K are  $ 1.47 \times 10^{-10}$ s$^{-1}$, $1.86 \times 10^{-11}$ s$^{-1}$, $3.37 \times 10^{-13}$ s$^{-1}$, $3.24 \times 10^{-11}$ s$^{-1}$ and $1.38 \times 10^{-10}$ s$^{-1}$, versus $1.38 \times 10^{-10}$ s$^{-1}$, $2.18 \times 10^{-11}$ s$^{-1}$, $5.31 \times 10^{-13}$ s$^{-1}$, $2.88 \times 10^{-11}$ s$^{-1}$ and $1.17 \times 10^{-10}$ s$^{-1}$ reported in the database.

\begin{table*}
\centering
\caption{Photodissociation rates for blackbody temperatures of HF and DF at 100 K, 1500 K, 3000 K, 5000 K and 10~000 K.}
\label{tab:rateshf}
\begin{tabular}{lcccccr} 
\hline
Blackbody field &	Isotopologue & $k$(100 K) (s$^{-1}$) & $k$(1500 K) (s$^{-1}$) & $k$(3000 K) (s$^{-1}$) & $k$(5000 K) (s$^{-1}$) & $k$(10000 K) (s$^{-1}$)) \\
\hline
 4000 K   & HF&$3.40\times 10^{-13}$&$4.82\times 10^{-13}$&$4.76\times 10^{-12}$&$2.84\times 10^{-10}$&$1.94\times 10^{-8}$  \\
                    & DF&$5.71\times 10^{-14}$&$8.82\times 10^{-14}$&$1.02\times 10^{-12}$&$1.20\times 10^{-10}$&$1.19\times 10^{-8}$  \\
\hline
 10~000 K & HF&$3.25\times 10^{-11}$&$3.38\times 10^{-11}$&$3.83\times 10^{-11}$&$5.67\times 10^{-11}$&$4.00\times 10^{-10}$  \\
                    & DF&$1.97\times 10^{-11}$&$2.07\times 10^{-11}$&$2.41\times 10^{-11}$&$3.86\times 10^{-11}$&$3.37\times 10^{-10}$   \\
\hline
 20~000 K & HF&$1.38\times 10^{-10}$&$1.40\times 10^{-10}$&$1.49\times 10^{-10}$&$1.76\times 10^{-10}$&$3.41\times 10^{-10}$  \\
                    & DF&$1.15\times 10^{-10}$&$1.18\times 10^{-10}$&$1.27\times 10^{-10}$&$1.52\times 10^{-10}$&$3.06\times 10^{-10}$  \\
\hline
\end{tabular}
\end{table*}

\section{Data provision}

A new data type called photodissociation cross sections (cm$^2$ molecule$^{-1}$)  has been created in the  \textsc{ExoMol} database, \href{www.exomol.com}{www.exomol.com} \citep{jt631}, based on the ExoMol data structure described by \citet{jt542}. For each isotopologue, a new entry will be added in the definition file supplementing the states and transitions file structure used to present line lists. In accordance with the ExoMol convention, the photodissociation data are not scaled with the isotopic abundance.  The
photodissociation cross sections will be published with low resolutions (1.0 nm) and high resolution (0.1 nm) spacing for the temperatures given in Table \ref{tab:temperatures}; results
for the present calculations will cover wavelengths  between 100 nm and 400 nm  with no pressure effects considered. 
Cross sections are stored in form of a space separated variable file. The file is structured as follows; the first column shows the wavelength, the second column reports the total cross section. The different temperature contributions are separated by a blank line. Table \ref{tab:database} shows as example the first four lines  of H$^{35}$Cl($T=0$ K) for the 1 nm spacing, the last two lines of H$^{35}$Cl($T=0$ K) with the first four lines of H$^{35}$Cl($T=100$ K). The line above the data shows the data formatting. The total and  partial cross sections contributions are reported in the electronic supporting information, with a small sample reported in Table \ref{tab:ESI}.

\begin{table}
\caption{Header of photodissociation cross section file for H$^{35}$Cl at $T=0$ K for the 1 nm spacing, also shown are the last two lines for $T = 0$ K and the first four lines for $T=100$ K. Wavelengths ($\lambda$) are in nm, and cross sections ($\sigma$) in \xsec. The first line gives the formatting used to write the data.}
\label{tab:database}
\centering
\begin{tabular}{ccccccccccc} 
\hline
$\lambda$ & $\sigma^{\textrm{Total}}$\\
\hline
F6.2 & es13.5\\
\hline
 100.00 &  4.85536E-20 \\
 101.00 &  1.11513E-19 \\
 102.00 &  1.37037E-18 \\
 103.00 &  4.82263E-19 \\
 104.00 &  1.21327E-18 \\
 $\vdots$ & $\vdots$  \\
 399.00 &  0.00000E+00 \\
 400.00 &  0.00000E+00 \\
  &      \\
 100.00 &  5.06233E-20 \\
 101.00 &  1.10736E-19 \\
 102.00 &  1.36532E-18 \\
 103.00 &  4.99004E-19 \\
 104.00 &  1.18854E-18 \\
\hline
\end{tabular}
\end{table}

\begin{landscape}
\begin{table}
\caption{Header of the photodissociation cross section table presented in the Supporting Information for H$^{35}$Cl for $T=0$ K for the 1 nm spacing, plus the last two lines for $T=0$ K and the first four for $T=100$ K. Wavelengths ($\lambda$) are in nm, and cross sections ($\sigma$) in \xsec. The first line reports the formatting used for writing the data.}
\label{tab:ESI}
\centering
\resizebox{1.3\textwidth}{!}{
\begin{tabular}{ccccccccccc} 
\hline
$\lambda$ & $\sigma^{\textrm{Total}}$ & $\sigma^{\textrm{A}\,^1\Pi\leftarrow \textrm{X}\,^1\Sigma^+}$ &  $\sigma^{\textrm{C}\,^1\Pi\leftarrow \textrm{X}\,^1\Sigma^+}$ & $\sigma^{\textrm{B}\,^1\Sigma^+\leftarrow \textrm{X}\,^1\Sigma^+}$ & $\sigma^{\textrm{D}\,^1\Pi\leftarrow \textrm{X}\,^1\Sigma^+}$ & $\sigma^{\textrm{H}^1\Sigma^+\leftarrow \textrm{X}\,^1\Sigma^+}$ & $\sigma^{\textrm{K}\,^1\Pi\leftarrow \textrm{X}\,^1\Sigma^+}$ & $\sigma^{\textrm{5}\,^1\Pi\leftarrow \textrm{X}\,^1\Sigma^+}$ & $\sigma^{\textrm{M}\,^1\Pi\leftarrow \textrm{X}\,^1\Sigma^+}$ & $\sigma^{\textrm{4}\,^1\Sigma^+\leftarrow \textrm{X}\,^1\Sigma^+}$ \\
\hline
1x,F6.2 & 10 $\times$(1x,es13.5)&&&&&&&&&\\
\hline
 100.00 &  4.85536E-20 &  2.49720E-25 &  7.385653E-23  &  1.438835E-21  &  1.16735E-22  &  1.37962E-21  &  2.14801E-22  & 1.96018E-22 &  2.35598E-23  &  4.51099E-20 \\
 101.00 &  1.11513E-19 &  7.06771E-25 &  1.918758E-22  &  2.143218E-21  &  1.83478E-22  &  5.26944E-21  &  9.95494E-22  &  1.75402E-21  & 1.74545E-21  &  9.92297E-20\\
 102.00 &  1.37037E-18 &  1.86128E-24 &  4.206999E-22  &  3.240703E-21  &  5.461986E-23  &  3.33646E-21  &  1.76091E-22  &  8.89503E-22 &  8.28554E-23  & 1.36217E-18\\
 103.00 &  4.82263E-19  & 4.52986E-24 &  3.515567E-22  &  4.953732E-21  &  1.11922E-23  &  5.20284E-21  &  7.70287E-22  &  2.39150E-21 &  2.65150E-20  &  4.42062E-19\\
 104.00 &  1.21327E-18 &  1.04052E-23 &  5.363629E-22  &  7.578607E-21 &  2.68010E-22 &  9.32299E-22 &  4.91039E-22 &  3.03492E-21 & 1.13187E-21  & 1.19929E-18\\
  & & & & &$\vdots$ & & & & & \\
 399.00 &  0.00000E+00 &  0.00000E+00 &  0.000000E+00 &  0.000000E+00 &  0.00000E+00 &  0.000000E+00 &  0.00000E+00  & 0.00000E+00  & 0.000000E+00 &  0.000000E+00\\
 400.00 &  0.00000E+00 &  0.00000E+00 &  0.000000E+00 &  0.000000E+00 &  0.00000E+00 &  0.000000E+00 &  0.00000E+00  & 0.00000E+00  & 0.000000E+00 &  0.000000E+00\\
  & & & & & & & & & & \\
 100.00 &  5.06233E-20  & 2.47899E-25 &  7.34458E-23 &  1.44117E-21 &  1.22165E-22 &  1.415554E-21 &  2.14494E-22  & 2.01564E-22 &  2.24269E-23 &  4.71323E-20\\
 101.00 &  1.10736E-19  & 7.02051E-25 &  1.97132E-22 &  2.14565E-21 &  1.78191E-22 &  5.281147E-21 &  9.81729E-22  & 1.73214E-21  & 1.76048E-21 &  9.84586E-20\\
 102.00 &  1.36532E-18  & 1.84883E-24 &  4.16114E-22 &  3.24302E-21 &  5.79011E-23 &  3.281736E-21 &  1.83332E-22  & 9.18835E-22 &  8.15775E-23 &  1.35714E-18\\
 103.00 &  4.99004E-19  & 4.49979E-24 &  3.48161E-22 &  4.95397E-21 &  1.36449E-23 &  5.242623E-21 &  7.99597E-22  & 2.32904E-21 &  2.62712E-20 &  4.59041E-19\\
 104.00 &  1.18854E-18  & 1.03338E-23 &  5.43194E-22 &  7.57846E-21 &  2.57864E-22 &  1.063908E-21 &  5.15449E-22  & 3.05713E-21 &  1.25755E-21 &  1.17425E-18\\
\hline
\end{tabular}
}
\end{table}
\end{landscape}

It is relatively easy to generate photodissociation rates as a function of molecular temperature  from the tabulated cross sections using a specified radiation field and Eq.~\ref{eq:radfield}.
It is our plan to implement an automated procedure to do this alongside a library of appropriate radiation fields.

\section{Conclusions}

In this paper we present the first in a series of papers investigating the temperature dependence of the photodissociation cross sections and rates for molecules of importance to exoplanetary, and of course other, studies. In this work we focus on two hydrogen halides, HCl and HF, and their isotopologues.

In case of HCl, we have considered 9 singlet electronic bands from the ground electronic state of the four main isotopes H$^{35}$Cl, D$^{35}$Cl, H$^{37}$Cl, and D$^{37}$Cl. We observe that an increase  in temperature leads to  the loss of details of bound-bound transitions and an increase in the cross sections at high wavenumber, given by the population of higher rotational and vibrational states in the electronic ground state. We calculate the photodissociation rates  for 5 different stellar fields, corresponding to the ISRF, Solar spectrum and three different blackbody fields. We observe that the dissociation rate in stellar fields corresponding to low temperature stars shows a large dependence on the molecular temperature, of at least two orders of magnitudes; the hot temperature fields instead are less sensitive to the temperature. The isotopic effect is prominent in the case of DCl, that, in case of 4000 K blackbody radiation field, shows a variation of the photodissociation rate of  two/one order of magnitude at $T> 3000$ K. 

We studied the electronic photodissociation spectrum between the electronic ground state and the first 3 electronic excited states of HF and DF. The photoabsorption cross sections show similar characteristics to HCl, but with the electronic states more separated than HCl. The photodissociation rates for the ISRF and  blackbody fields at 10~000 K and 20~000 K show the same temperature trend as HCl, while the blackbody field of 4000 K shows  completely different behaviour at low temperatures with respect to the Solar field: the rates obtained using the  blackbody of 4000 K  are two orders of magnitude lower than the rates obtained by the Solar field. 

The rapid rise in photodissociation rates with temperature in typical radiation fields of cool stars is important. This process is largely driven by the fact that in general vibrational excitation leads to a lowering of the photodissociation threshold which is typically several times larger than the vibrational excitation energy involved, see \citet{jt229} for a discussion of this. HF and HCl actually have quite large vibrational quanta meaning that thermal vibrational excitation only becomes significant at temperatures  over 3000 K. For most other
molecules of importance for exoplanetary atmospheres one would expect the sharp rise in photodissociation rates to start at lower temperatures due to their smaller spacing between vibrational levels. However, we show that, particularly for cool stars, the precise form of the stellar radiation field used in any model may be 
crucial in obtaining reliable results. This finding, together with the recent work from \citet{22TeKeBa.fields}, put emphasis on the need to correctly characterise these fields.

The discussion of rates in this paper is based entirely on the assumption that the molecule concerned is thermalised. However, non-local thermodynamic equilibrium (NLTE) effects are thought to be important in exoplanets. It would be relatively straightforward to adapt our procedure to provide NLTE data, but this would require re-summing individual vibrational and rotational contributions to the cross sections.

Our photodissociation models for HCl and HF is based on two assumptions. The first assumption is that every excitation leads to a dissociative event, as discussed in the methods section. 
The second assumption concerns  the use of the adiabatic B~$^1\Sigma^+$ for HCl instead of the diabatic  V~$^1\Sigma^+$ ionic  and E~$^1\Sigma^+$ Rydberg states. While this choice is justified in case  of photodissociation, it may not be optimal for building a spectroscopic model, as shown by the high resolution REMPI spectra of \citet{91GrBiWa.HCl}.  In this case, the rovibrational levels spacing of the two states become an important factor to consider. The  E~$^1\Sigma^+$ rotational constants are comparable to the X~$^1\Sigma^+$ ones, while the V~$^1\Sigma^+$ rotational constants are smaller by a factor of two \citep{81GiGixx.HCl,90BeKoKo.HCl,91GrBiWa.HCl,13LoWaKv.HCl}. Similarly, to obtain details of the product distribution it is necessary to consider the singlet-triplet couplings. The triplet states (a~$^3\Pi$, b~$^3\Pi$, t~$^3\Sigma^+$, \textit{etc}...)  are important for  final product states analysis  \citep{98AlLiLi.HCl,99ReLaCo.HCl,00ReAsBr.HCl}. Implementation of model
which includes triplet states and the appropriate spin-orbit couplings would allow 
 final state analysis of photodissociation processes to be performed which would improve our understanding of these phenomena, and it will help in the understanding of the tiplet excitation contributions for  vibrationally excited molecules, quantifying their overall contributions for high temperatures. This  problem is left to future work.

\section*{Data availability}

\textsc{Duo} and \textsc{ExoCross} input files  HCl and HF as used for generating the photodissociation cross sections and rates are available as supplementary information in archives \textit{HCl.zip} and \textit{HF.zip}. The open access programs \textsc{Duo} and \textsc{ExoCross} are available from  github.com/exomol. 
 
The computed photodissociation cross sections are available from the ExoMol website, www.exomol.com and as supplementary material to this paper.

\section*{Acknowledgments}

We thank Olivia Venot for helpful discussions over the course of this work.  This work was funded by ERC Advanced Investigator Project 883830 and by the STFC Project  ST/R000476/1.


\bibliographystyle{mnras}
\bibliography{journals_astro,HCl,HF,jtj,methods,photodissociation,radfield,exoplanets,non-LTE,Books} 


\end{document}